\documentclass[8.5pt,twoside,twocolumn]{article}
\usepackage[super,sort&compress,comma]{natbib}
\usepackage{mhchem}
\usepackage{times,mathptmx}
\usepackage{sectsty}
\usepackage{balance}

\usepackage{graphicx} 
\usepackage{lastpage}
\usepackage[format=plain,justification=raggedright,singlelinecheck=false,font=small,labelfont=bf,labelsep=space]{caption}
\usepackage{fancyhdr}
\begin{document}

\thispagestyle{plain}
\fancypagestyle{plain}{
\renewcommand{\headrulewidth}{1pt}}
\renewcommand{\thefootnote}{\fnsymbol{footnote}}
\renewcommand\footnoterule{\vspace*{1pt}%
\hrule width 3.4in height 0.4pt \vspace*{5pt}}
\setcounter{secnumdepth}{5}

\makeatletter
\def\subsubsection{\@startsection{subsubsection}{3}{10pt}{-1.25ex plus -1ex minus -.1ex}{0ex plus 0ex}{\normalsize\bf}}
\def\paragraph{\@startsection{paragraph}{4}{10pt}{-1.25ex plus -1ex minus -.1ex}{0ex plus 0ex}{\normalsize\textit}}
\renewcommand\@biblabel[1]{#1}
\renewcommand\@makefntext[1]%
{\noindent\makebox[0pt][r]{\@thefnmark\,}#1}
\makeatother
\renewcommand{\figurename}{\small{Fig.}~}
\sectionfont{\large}
\subsectionfont{\normalsize}

\fancyfoot{}
\fancyfoot[RO]{\footnotesize{\sffamily{1--\pageref{LastPage} ~\textbar  \hspace{2pt}\thepage}}}
\fancyfoot[LE]{\footnotesize{\sffamily{\thepage~\textbar\hspace{3.45cm} 1--\pageref{LastPage}}}}
\fancyhead{}
\renewcommand{\headrulewidth}{1pt}
\renewcommand{\footrulewidth}{1pt}
\setlength{\arrayrulewidth}{1pt}
\setlength{\columnsep}{6.5mm}
\setlength\bibsep{1pt}

\twocolumn[
  \begin{@twocolumnfalse}
\noindent\LARGE{\textbf{Fingering to fracturing transition in a transient gel}}
\vspace{0.6cm}

\noindent\large{\textbf{Guillaume Foyart,\textit{$^{a,b}$} Laurence Ramos$^{\ast}$\textit{$^{a,b}$}, Serge Mora,\textit{$^{a,b}$}
and Christian Ligoure$^{\ast}$\textit{$^{a,b}$}
}}\vspace{0.5cm}

\noindent\textit{\small{\textbf{Received Xth XXXXXXXXXX 20XX, Accepted Xth XXXXXXXXX 20XX\newline
First published on the web Xth XXXXXXXXXX 200X}}}

\noindent \textbf{\small{DOI: 10.1039/b000000x}}
\vspace{0.6cm}

\noindent \normalsize{Fracture processes are ubiquitous in soft materials, even in complex fluids, subjected to stresses. To investigate these processes in a simple geometry, we use a model self-assembled transient gel and study the instability patterns obtained in a radial Hele-Shaw cell when a low viscosity oil pushes the more viscous transient gel. Thanks to an analysis of the morphology of the patterns, we find a discontinuous transition between the standard Saffman-Taylor fingering instability and a fracturing instability as the oil injection rate increases. Our data suggest that the flow properties of the gel ahead of the finger tip controls the transition towards fracturing. By analyzing the displacement field of the gel in the vicinity of the fingers and cracks, we show that in the fingering regime, the oil gently pushes the gel, whereas in the fracturing regime, the crack tears apart the gel, resulting in a strong drop of the gel velocity ahead of the crack tip as compared to the tip velocity. We find a unique behavior for the whole displacement field of a gel around a crack, which is drastically different from that around a finger, and reveals the solid-like behavior of the gel at short time. Our experiments and analysis provide quantitative yet simple tools to unambiguously discriminate a finger from a crack in a visco-elastic material.
 }
\vspace{0.5cm}
\end{@twocolumnfalse}
]

\section{Introduction}


\footnotetext{\textit{$^{a}$~Universit\'{e} Montpellier 2, Laboratoire Charles Coulomb UMR 5221, F-34095, Montpellier,
France.}}
\footnotetext{\textit{$^{b}$~CNRS, Laboratoire Charles Coulomb UMR 5221, F-34095, Montpellier, France.}}
\footnotetext{\textit{$^{*}$~E-mail: laurence.ramos@univ-montp2.fr; christian.ligoure@univ-montp2.fr}}


When submitted to mechanical stresses, a large variety of soft materials, comprising complex fluids as colloidal suspensions and transient networks,  may flow as fluids or fracture as solids, resulting from the delicate balance between the viscous and elastic components of the sample visco-elasticity. Several geometries (see the review \cite {Ligoure2013} and references therein) have been considered to explore fracture mechanisms in soft materials. These include impacting the soft material with a solid, imposing an elongational flow using either an extensional rheometer or thanks to a pendant drop experiment, and imposing a shear flow in a rheometric cell. Forcing the soft material to flow in a two-dimensional cell (Hele–Shaw cell) appears as a very simple configuration to investigate fracture processes in soft matter. This is the geometry chosen in this paper. Here, we analyze the instabilities patterns that result from the displacement of one fluid by another fluid of lower viscosity. Very generally, the instabilities patterns are of two kinds: when a Newtonian high-viscosity fluid is penetrated by a low-viscosity fluid, fingering develops at the interface between the two fluids as a result of the destabilizing action of viscous forces. This is the classical Saffman-Taylor instability \cite{Saffman1958,Bensimon1986}. By contrast, if the invaded fluid is visco-elastic, at high Deborah number, fracturing develops, as driven by the release of the elastic stress \cite{Ligoure2013}. Note in addition that elastic fingering instabilities have also been predicted and observed for Maxwell fluids \cite{Mora2009,Mora2012}. Fingering to fracturing transitions have been reported in several classes of complex fluids, including clay suspensions \cite {Lemaire1991}, lyotropic lamellar phases \cite{Greffier1998}, liquid foams \cite{Bensalem2013} and self-assembled transient networks \cite{Zhao1993,Ignes-Mullol1995,Vlad1999}. So far, the quantitative analysis of this transition was  mainly restricted to (i) the mass  fractal dimension of  the patterns which ranges from $1.4$ to $1.7$ for a viscous fingering pattern and drops to $\sim 1$ for crack patterns \cite{Lemaire1991,Zhao1993}, (ii) the tip velocity of the instability \cite{Ignes-Mullol1995,Vlad1999}, (iii) the nucleation of cracks in the particular case where the elastic modulus of the viscoelastic fluid is low enough that the growth rate of the Saffman-Taylor instability is of the same order of magnitude as the Maxwell relaxation time \cite{Mora2009,Mora2012}, allowing a limited understanding of the physical mechanisms at the origin of the transition. In particular quantifications of the morphology of the finger and crack tips and of the velocity field of the viscoelastic fluid in the vicinity of the interfaces are still missing.
In this article, using a transient gel \cite{Chassenieux2011,Ligoure2013} confined in a Hele-Shaw radial cell, we
will show that these parameters are essential to apprehend the transition. In particular, we highlight the morphology differences between fingers and cracks and demonstrate that the gel velocity in the direction of the crack motion is strongly reduced compared to that of the crack tip, whereas a continuity of the velocity at the oil/gel interface is found for fingers, thus providing quantitative yet simple tools to unambiguously discriminate a finger from a crack.

\section{Materials and methods}
\label{sec:methods}

The experimental system is a self-assembled transient gel consisting in an aqueous suspension of surfactant-stabilized oil-droplets, of diameter $6.2$ nm, reversibly linked by telechelic polymers. Both at rest and under flow, the droplets are spherical and isotropically distributed in the aqueous solvent. The sample composition and viscoelasticity have been described elsewhere \cite{Michel2000}. In brief, we use a mixture of cetylpyridinium chloride and n-octanol as resp. surfactant and cosurfactant, decane as oil and a $0.2$ M NaCl solution as aqueous solvent. The telechelic polymer is a home-made triblock copolymer, comprising a central polyethylene oxide water-soluble block (of molecular weight $10000$ g/mol) at the extremity of which hydrophobic aliphatic stickers of $18$ carbons each are covalently bound. The droplets mass fraction is $10 \%$ and the molar ratio of polymer over surfactant+cosurfactant is 0.00384, yielding on average $10$ stickers per droplet. The oil is eventually colored with red Sudan IV to increase the contrast between the oil and the gel and the gel is eventually seeded with a small amount ($1$wt $\%$) of silica particles of average size $50 \, \mu$m (silica gel $60$ by Merck) for image correlation velocimetry measurements. We have checked that the addition of particles does not modify significantly the sample viscoelasticity. The transient gel is a Maxwell fluid with a shear plateau modulus, $G_0 = (1650 \pm 50) $ Pa, and a characteristic relaxation time, $\tau = (0.70 \pm 0.03)$ s. The flow curve, shear stress $\sigma$ versus shear rate $\dot{\gamma}$, of this gel has been characterized \cite{Tabuteau2009}. For $ \dot{\gamma} \leq 1/ \tau$, the steady shear stress is proportional to the shear rate, as for a Newtonian fluid with viscosity $\eta = G_0 \times \tau$. When $\dot{\gamma}$ reaches $1/ \tau$,  an abrupt drop of $\sigma$ occurs, which has been unambiguously identified as the signature of a  shear-induced fracture in the material  \cite{Tabuteau2009}.

The experimental set-up is a standard radial Hele-Shaw cell, where the gel is confined between two glass plates (thickness $5$ mm) separated by $500 \, \mu$m thick Mylar spacers.  Colza oil of zero-shear viscosity $60$ mPa.s is injected at a constant volume rate, $Q$, using a syringe pump through a $4$ mm hole drilled in one of the two plates, and pushes the gel. In our experiments, $Q$ is varied between $0.01$ and $40$ ml/min. The visualization of the whole cell and of the oil-gel interface is achieved with a fast CMOS camera (Phantom v7), usually used at an acquisition rate of $500$ Hz.

\begin{figure}[h]
\centering
  \includegraphics[height=6.5cm]{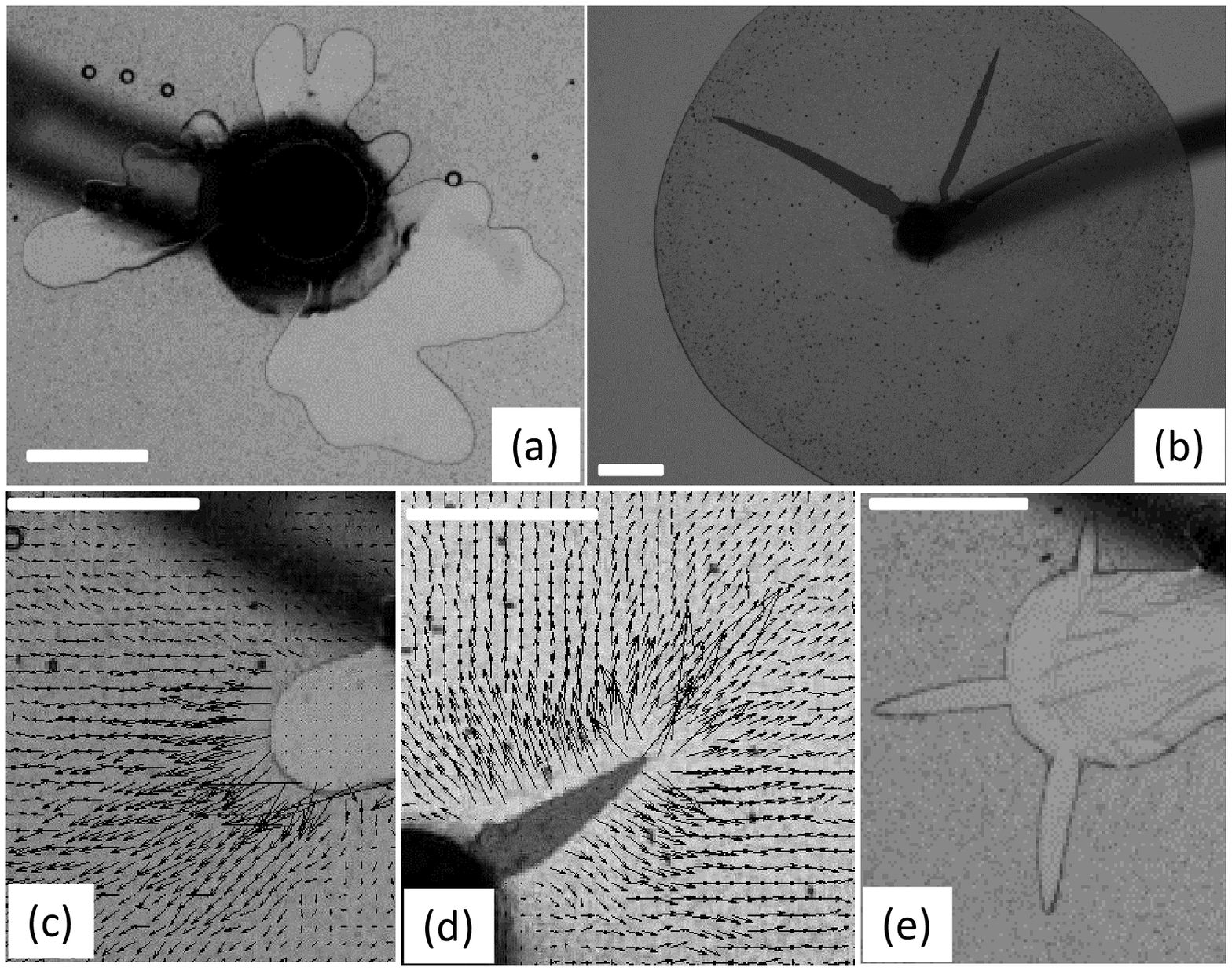}
  \caption{ Visualization of the oil-gel interface when the oil is injected at different rates, $Q$, and pushes the gel. (a) $Q = 0.04$ ml/min, (b) $Q = 2$ ml/min; (c, d, e) Zoom up of interface for $Q = 0.04$ ml/min (c), $Q = 1$ ml/min (d) and when a jump from $0.04$ to $10$ ml/min is performed (e). The arrows in (c,d) show the displacement field in the gel, as determined by image correlation velocimetry. Scale bars: $6$  mm.}
  \label{fig:1}
\end{figure}

\section{Results and Discussions}
\label{sec:resdisc}

\begin{figure}[h]
\centering
  \includegraphics[height=10cm]{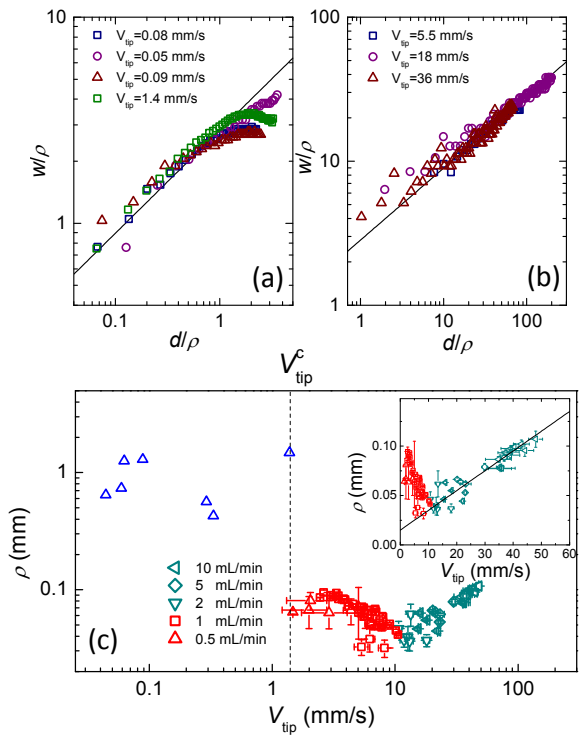}
  \caption{Color online. Width, $w$, of the instability as a function of the distance from the tip of the instability, $d$. $w$ and $d$ are normalized by the radius of curvature $\rho$ of the pattern. Results for fingers and cracks are shown respectively in (a) and (b), for instabilities propagating at different velocities as indicated in the legend. The dotted line is the theoretical expectation for a parabolic shape. Radius of curvature of the fingers and cracks as a function of their speed of propagation. (Inset) Zoom of the data shown in the main plot for cracks.}
  \label{fig:2}
\end{figure}

At low injection rates, the gel behaves as a Newtonian fluid, and the instability of the oil-gel interface observed when the low-viscosity oil pushes the more viscous gel is the classical Saffman-Taylor finger instability \cite{Bensimon1986} (fig.~\ref{fig:1}a,c).  When the injection rate increases, the oil-gel interface instability leads to a markedly different morphology. As shown in fig.~\ref{fig:1}a,c, the finger instabilities obtained at low $Q$ are rather fat and roundish whereas thinner and sharper instability patterns are obtained for higher $Q$  (fig.~\ref{fig:1}b, d). The sharp contrast between the two patterns is nicely stressed in fig.~\ref{fig:1}e, thanks to a jump from a low ($Q = 0.04$ ml/min) to a high ($Q = 10$ ml/min) injection rates. To quantify the shape of the instability patterns, we measure the width of the pattern, $w$, as a function of the distance from the tip, $d$, fit the profile with a parabola and extract a radius of curvature, $\rho$. To allow for a comparison of different patterns, the two lengths, $w$ and $d$, are normalized by $\rho$. Results are plotted in fig.~\ref{fig:2}. For the finger pattern, one measures a parabolic shape, $w/\rho =  2 \sqrt{2d/\rho}$ up to a distance from the tip equal to the radius of curvature.  Significant departures from a parabolic shape are systematically measured at longer distances, as shown in fig.~\ref{fig:2}a. In addition, the non-universal behavior for $d > \rho$ presumably reflects the interaction between the different fingers due to the radial geometry.  For the sharper instability patterns obtained for higher $q$, no data are available for $d< \rho$ due to the resolution of our set-up ($1$ pixel $\simeq 0.09$ mm), but we interestingly find that the shape remains parabolic up to distances more than $100$ times the measured radius of curvature (fig.~\ref{fig:2}b). The parabolic shape over long distances suggests that this type of instability pattern can be considered as a crack, as observed for brittle solids \cite{Irwin1957,Saulnier2004,Tabuteau2011}.  Note that we find an overlap of the data for cracks with different radii of curvature spanning one order of magnitude, suggesting a universality of the shape. These cracks were generated using different injection rates spanning almost two orders of magnitude. They propagate at different velocities, $V_{\rm{tip}}$, as determined by tracking the crack tip by image analysis.  Note that $V_{\rm{tip}}$ is the instantaneous tip velocity that may vary with time and depends, because of volume conservation, on the number and shape of the fingers or cracks and on the imposed flow rate. The radius of curvature is plotted as a function of $V_{\rm{tip}}$ in fig.~\ref{fig:2}c. The plot clearly shows a sharp and discontinuous transition at $V^c_{\rm{tip}}= (1.4 \pm 0.1)$ mm/s between a fingering regime, when $V_{\rm{tip}}<V_{\rm{tip}}^c$, to a fracturing regime, when $V_{\rm{tip}}>V_{\rm{tip}}^c$. In the fingering regime, one finds that the radius of curvature of the finger fluctuates around an average value $\rho = (0.92 \pm 0.52)$ mm. By contrast, in the fracturing regime, the radius of curvature is about one order of magnitude smaller ($0.01 < \rho < 0.1$ mm) and displays a non-monotonic evolution with the tip velocity (as shown in the inset of fig.~\ref{fig:2}c).

Although the exact way the gel flows in the thin Hele-Shaw cell when pushed by the oil is not known, from a dimensional analysis, one expects the gel to be sheared at an average shear rate $\dot{\gamma} \simeq \frac{V_{\rm{tip}}}{b}$, where $b=0.5$ mm is the gap of the cell. One therefore finds that, when a finger moves at the critical velocity, $V_{\rm{tip}}^c$, the shear rate experienced by the gel is of the order of $2.8 \, \rm{s}^{-1}$. Interestingly, this numerical value is comparable to the critical shear rate $1/ \tau = 1.4 \, \rm{s}^{-1}$ at which fracture occurs in a shear experiment, although higher, as expected since the Hele-Shaw experiments are non-steady experiments as opposed to the shear experiments where the critical shear rate correspond to measurements obtained in steady state.

\begin{figure}[h]
\centering
  \includegraphics[height=3.8cm]{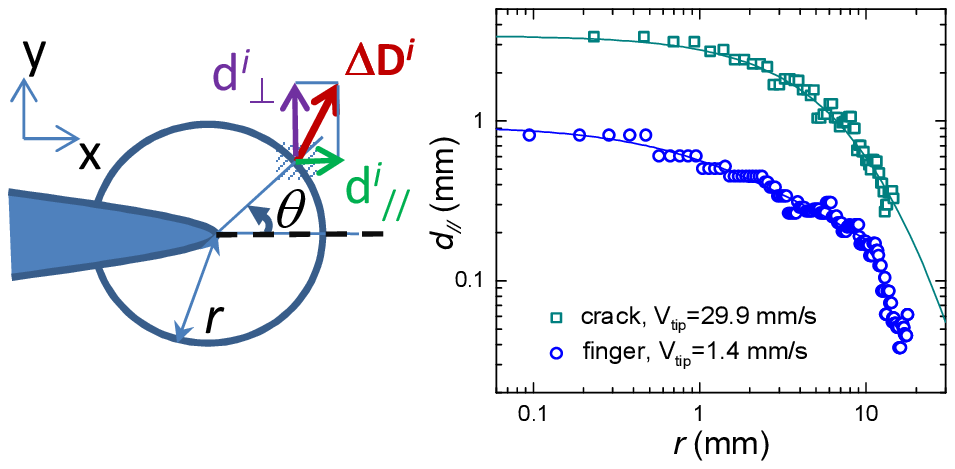}
  \caption{
 Color online. (left) Scheme showing the definitions of angle and axis. (right) Plot of $d_{\parallel}$ as a function of $r$ the distance from the tip for data acquired along the dotted line shown in the scheme. Lines are best fits of the data shown as symbols, for a finger (blue circles) and a crack (green squares), used to evaluate the extrapolation towards $r=0$.}
\label{fig:3}
\end{figure}

\begin{figure}[h]
\centering
  \includegraphics[height=10cm]{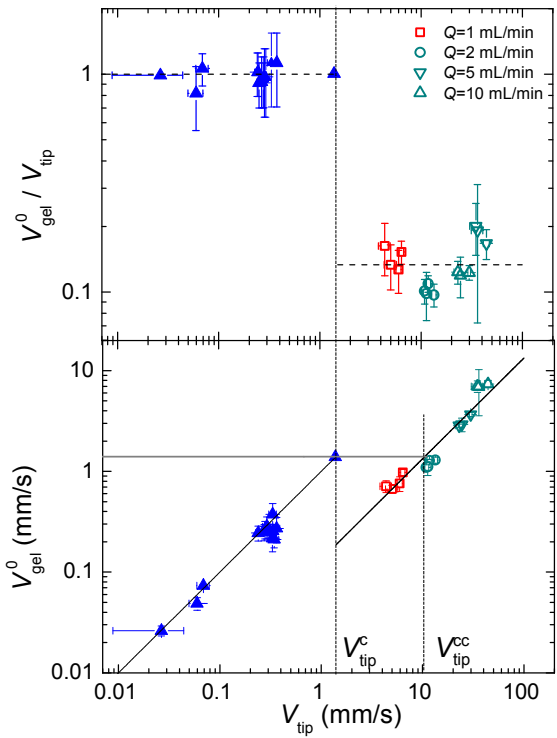}
  \caption{Color online. (a) Ratio between the velocity of the gel at the oil/gel interface ahead of the finger or crack,  $V^0_{\rm{gel}}$, and the velocity of the finger or crack, $V_{\rm{tip}}$, as a function of $V_{\rm{tip}}$. (b) Evolution of $V^0_{\rm{gel}}$ with $V_{\rm{tip}}$ for fingers and cracks.}
    \label{fig:4}
\end{figure}

We have shown that the morphology of the instability pattern can be used as a criterion to discriminate a crack from a finger. In the following, we investigate the motion of the gel in the vicinity of the oil-gel interface and demonstrate that our analysis provides an alternative quantitative tool to allow one to analyze the finger to crack transition. Image correlation velocimetry was performed using standard procedure for samples seeded with particles to quantify the displacement field. Each image is divided into regions of interest (ROIs) of $10$ pixels $\times 10$ pixels size (corresponding to $0.86 \times 0.86 \rm{mm}^2$), and the translational motion of each ROI $i$, $ \Delta \textbf{D}^i = d^i_{\parallel} \textbf{x} + d^i_{\perp} \textbf{y} $ for pairs of images taken at times $t$ and $t+\delta t$ is measured. We define as $\textbf{x}$  (resp. $\textbf{y}$) the direction parallel (resp. perpendicular) to the finger or crack propagation direction between times $t$ and $t+\delta t$ ; $d^i_{\parallel}$ (resp. $d^i_{\perp}$) is the displacement of ROI $i$ along the $\textbf{x}$ (resp. $\textbf{y}$) direction (see scheme fig.~\ref{fig:3}). In a first step, we investigate how the gel is displaced ahead of the finger or crack. To do so, we focus on $d_{\parallel}$, the displacement field in the direction parallel to the displacement of the tip, and we measure the evolution of $d^i_{\parallel}$ with the distance $r$ from the tip, for a line of ROIs parallel to the direction of propagation (thick dotted line in the scheme fig.~\ref{fig:3}). Examples of the dependence of $d^i_{\parallel}$  with $r$ for a finger and a crack are provided in the plot fig.~\ref{fig:3}. In both cases, $d_{\parallel}(r)$ increases when $r$ decreases but saturates for vanishingly small $r$. This allows one to evaluate the instantaneous gel velocity at the front tip, $V^{0}_{\rm{gel}}$ as $d_{\parallel}(r \rightarrow 0) / \delta t$, where $d_{\parallel}(r \rightarrow 0)$ is the extrapolation of $d_{\parallel}(r)$ at $r=0$ (obtained by fitting  $d_{\parallel}(r)$ with the functional form, $\frac{a}{(r+b)^n}$, where the plateau when $r \rightarrow 0$ is $ab^{-n}$) and $\delta t$ is the delay between the two images used to calculate the displacement field. For fingers, we find that the instantaneous gel velocity is within errors equal to the tip velocity ($V^0_{\rm{gel}} / V_{\rm{tip}} = 0.98 \pm 0.09 $) as expected. Interestingly, for cracks one systematically measures that $V^{0}_{\rm{gel}}$ is smaller that $V_{\rm{tip}}$: $V^0_{\rm{gel}} / V_{\rm{tip}} = 0.14 \pm 0.03 $. The measurements are gathered in fig.~\ref{fig:4}a where the ratio $V^0_{\rm{gel}} / V_{\rm{tip}}$ is plotted as a function of $V_{\rm{tip}}$. Again, this plot demonstrates a sharp and discontinuous transition between a fingering regime, when $V_{\rm{tip}} < V^{\rm {c}}_{\rm{tip}}$ for which  $V^0_{\rm{gel}} / V_{\rm{tip}} \simeq 1$ and a fracturing regime, when $V_{\rm{tip}} > V^{\rm {c}}_{\rm{tip}}$,  where $V^0_{\rm{gel}} / V_{\rm{tip}} \ll 1$.

It is also instructive to re-examine the data by plotting $V^0_{\rm{gel}}$ as a function of $V_{\rm{tip}}$ (fig.~\ref{fig:4}b).
By doing so, we can identify two regions, depending on $V^0_{\rm{gel}}$ being higher or lower than $V^{\rm {c}}_{\rm{tip}}$. For $ V_{\rm{tip}}> V^{\rm{cc}}_{\rm{tip}} \simeq 10$ mm/s, $V^0_{\rm{gel}} > V^{\rm{c}}_{\rm{tip}}$. Hence at the crack tip, the gel is on average sheared at a shear rate larger than the inverse of its characteristic relaxation time. It can thus be considered as an elastic material. Generally, for brittle cracks propagating in an elastic medium, the fracture energy, $J$, can be related to the radius of curvature of the crack \cite{Hui2003,Tabuteau2011}.
For an incompressible elastic medium with a shear modulus $G_0$ in
a plain stress geometry, $J=\frac{3}{4} \frac{\pi G_0 \rho}{2}$. Our data show therefore that the fracture energy linearly increases with $V_{\rm{tip}}$ as does the radius of curvature of the crack (inset of fig.~\ref{fig:2}c). This is similar to what has been measured for gelatin \cite{Baumberger2006}. The extrapolation of $J$ towards $0$ velocity yields $J_0=(45\pm12) \rm{ mJ/m}^2$. This value should be equal to twice the interfacial tension between the gel and the oil. It is not possible to measure the surface tension between the oil and the gel. Instead, we measured the surface tension between the oil and a low viscosity microemulsion sample of composition equal to that of the gel but without the telechelic polymer. We found a surface tension of $29 \, \rm{ mJ/m}^2$, in good agreement with the theoretical expectation of $J_0/2$.

On the other hand, for  $V^{\rm{c}}_{\rm{tip}}<V_{\rm{tip}}< V^{\rm{cc}}_{\rm{tip}}$, $V^0_{\rm{gel}} < V^{\rm{c}}_{\rm{tip}}$,
 the tip velocity is small enough to lead to a velocity of the gel at the front of the tip smaller than $V^c_{\rm{tip}}$, suggesting that in this zone the viscous character of the gel plays a role. Interestingly and unexpectedly, one finds that in this region,  the radius of curvature of the crack decreases with the tip velocity, which presumably hints at some dissipation mechanism due to the visco-elasticity of the gel.

\begin{figure}[h]
\centering
  \includegraphics[height=7cm]{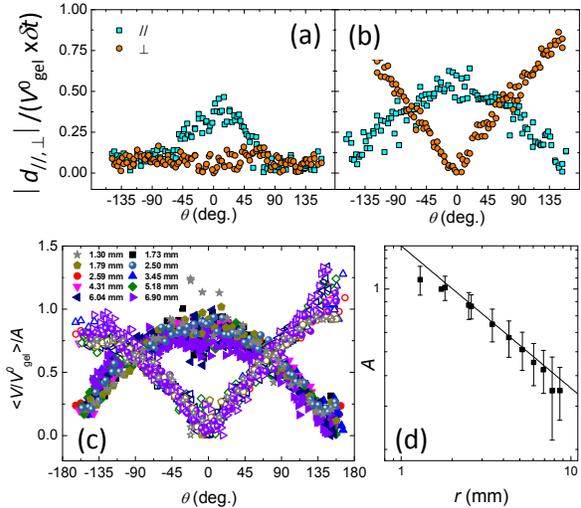}
  \caption{Color online.
Analysis of the displacement field around a finger (a) and a crack (b). Evolution of $|d_{\parallel} (r, \theta)| / (V_{\rm{tip}} \delta t)$ and $|d_{\perp} (r, \theta)| / (V^0_{\rm{gel}} \times \delta t)$ with angle $\theta$ for a given distance from the tip, $r$. In (a) $r=4$ mm and in (b) $r=3.75$ mm. (c) Master curve obtained for data acquired at different $r$ as indicated in the legend. (d) Prefactor used to obtained the master curve shown in (c). The symbols are the experimental data and the line is a power-law fit yielding an exponent $-0.58$.
}
\label{fig:5}
\end{figure}

So far, we have seen that, in the fingering regime, the fingers gently push the gel leading to a continuity of the velocity, whereas in the fracturing regime the cracks tear apart the gel resulting in a strong drop of the gel velocity ahead of the crack tip and $V^0_{\rm{gel}} / V_{\rm{tip}} \ll 1$. This implies that in the crack regime the whole displacement field of the gel in the vicinity of the crack is modified, compared to that around a finger. This can be visually inferred in fig.~\ref{fig:1}c,d when we have superimposed to the picture of a crack and finger the displacement field measured by image correlation velocimetry.  To quantify, one considers the origin as the crack or finger tip, and look at the displacement field as a function of the angle $\theta$ in a circle of radius $r$. Origin of $\theta$ is taken as the direction of the displacement of the finger or crack (see scheme fig.~\ref{fig:3}). The displacements are projected along the direction of propagation, $d_{\parallel} (r, \theta)$ and perpendicular to this direction $d_{\perp} (r, \theta)$.
The evolution at a fixed distance from the tip of $|d_{\parallel} (r, \theta)|$ and $|d_{\perp} (r, \theta)|$ with angle $\theta$ is shown in fig.~\ref{fig:5}a, for a finger and in fig.~\ref{fig:5}b for a crack, where the data have been normalized by $V^0_{\rm{gel}} \delta t$, the distance the gel ahead of the tip has been displaced during the lag time $\delta t$.  Whereas $|d_{\parallel} (r, \theta)|$ shows for both fingers and cracks a similar bell-like shape, drastically different behaviors are obtained for $|d_{\perp} (r, \theta)|$: for a finger, $|d_{\perp} (r, \theta)|$ is small as compared to the parallel component and almost $\theta$-independent. By contrast, for a crack, $|d_{\perp} (r, \theta)|$ displays a $V$-shape and reaches numerical values comparable to those in the parallel direction. This shape reflects the tearing of the gel in front of a propagating crack. Note that the shapes of the displacement fields for fingers and cracks are robust. Data obtained for a fixed distance $r$ but for cracks propagating at different velocities (with $V_{\rm{tip}}$ both smaller or higher than $V^{\rm{cc}}_{\rm{tip}}$) have been first normalized by $V^0_{\rm{gel}}$ and subsequently averaged. In a third step average data obtained for different $r$ are rescaled by a $r$-dependent prefactor, $A$. The rescaled data are found to fall on a unique master curve, as reported in fig.~\ref{fig:5}c. This unambiguously shows the universality of the displacement field of the gel around a crack. We find in addition that the prefactor, $A$, decays as $r^{-p}$ with $p \simeq 0.58 \pm 0.05$, consistent with the expected long-range decay of an elastic medium (fig.~\ref{fig:5}d) \cite{Rice1968}.

\section{Conclusion}

We have studied a self-assembled transient gel confined in a radial Hele-Shaw cell and pushed by a low-viscosity oil and have visualized the oil/gel interface with a high-speed camera. Image analysis has allowed us to highlight the sharp transition from a Saffman-Taylor fingering instability to a fracturing mechanism, when the finger tip velocity increases, as revealed by distinct morphologies of the instability patterns. Our measurements have suggested that the transition is governed by the flow of the gel close to the oil/gel interface. We have analyzed the displacement field of the gel in the vicinity of the oil/gel interface and have evidenced drastically different displacement fields for fingers and cracks.
Thus our experiments provide quantitative tools to experimentally characterize a fingering to fracturing transition in gels and show the interplay between the gel visco-elasticity and the way fingers and cracks propagate. Our approach could be used to a large variety of soft materials. In this framework, we are currently investigating more complex transient gels, with tunable structures, which exhibit a brittle to ductile transition, as their morphology varies \cite{Tixier2010,Ramos2011}.

\section*{Acknowledgments}

Discussions with E. Bouchaud, M. Ciccotti, L. Cipelletti and O. Dauchot are acknowledged. This work has been
supported by ANR under Contract No. ANR-2010-BLAN-0402-1
(F2F).

\footnotesize{
\bibliography{biblio-F2F}

\providecommand*{\mcitethebibliography}{\thebibliography}
\csname @ifundefined\endcsname{endmcitethebibliography}
{\let\endmcitethebibliography\endthebibliography}{}
\begin{mcitethebibliography}{22}
\providecommand*{\natexlab}[1]{#1}
\providecommand*{\mciteSetBstSublistMode}[1]{}
\providecommand*{\mciteSetBstMaxWidthForm}[2]{}
\providecommand*{\mciteBstWouldAddEndPuncttrue}
  {\def\EndOfBibitem{\unskip.}}
\providecommand*{\mciteBstWouldAddEndPunctfalse}
  {\let\EndOfBibitem\relax}
\providecommand*{\mciteSetBstMidEndSepPunct}[3]{}
\providecommand*{\mciteSetBstSublistLabelBeginEnd}[3]{}
\providecommand*{\EndOfBibitem}{}
\mciteSetBstSublistMode{f}
\mciteSetBstMaxWidthForm{subitem}
{(\emph{\alph{mcitesubitemcount}})}
\mciteSetBstSublistLabelBeginEnd{\mcitemaxwidthsubitemform\space}
{\relax}{\relax}

\bibitem[Ligoure and Mora(2013)]{Ligoure2013}
C.~Ligoure and S.~Mora, \emph{Rheol. Acta}, 2013, \textbf{52}, 91\relax
\mciteBstWouldAddEndPuncttrue
\mciteSetBstMidEndSepPunct{\mcitedefaultmidpunct}
{\mcitedefaultendpunct}{\mcitedefaultseppunct}\relax
\EndOfBibitem
\bibitem[Saffman and Taylor(1958)]{Saffman1958}
P.~G. Saffman and G.~I. Taylor, \emph{Proc. R. Soc. London, Ser. A}, 1958,
  \textbf{245}, 312\relax
\mciteBstWouldAddEndPuncttrue
\mciteSetBstMidEndSepPunct{\mcitedefaultmidpunct}
{\mcitedefaultendpunct}{\mcitedefaultseppunct}\relax
\EndOfBibitem
\bibitem[Bensimon \emph{et~al.}(1986)Bensimon, Kadanoff, Liang, Shraiman, and
  Tang]{Bensimon1986}
D.~Bensimon, L.~P. Kadanoff, S.~Liang, B.~I. Shraiman and C.~Tang, \emph{Rev.
  Modern. Phys}, 1986, \textbf{58}, 958\relax
\mciteBstWouldAddEndPuncttrue
\mciteSetBstMidEndSepPunct{\mcitedefaultmidpunct}
{\mcitedefaultendpunct}{\mcitedefaultseppunct}\relax
\EndOfBibitem
\bibitem[Mora and Manna(2009)]{Mora2009}
S.~Mora and M.~Manna, \emph{Phys. Rev. E}, 2009, \textbf{80}, 016308\relax
\mciteBstWouldAddEndPuncttrue
\mciteSetBstMidEndSepPunct{\mcitedefaultmidpunct}
{\mcitedefaultendpunct}{\mcitedefaultseppunct}\relax
\EndOfBibitem
\bibitem[Mora and Manna(2012)]{Mora2012}
S.~Mora and M.~Manna, \emph{Journal of Non-Newtonian Fluid Mechanics}, 2012,
  \textbf{173-174}, 30\relax
\mciteBstWouldAddEndPuncttrue
\mciteSetBstMidEndSepPunct{\mcitedefaultmidpunct}
{\mcitedefaultendpunct}{\mcitedefaultseppunct}\relax
\EndOfBibitem
\bibitem[Lemaire \emph{et~al.}(1991)Lemaire, Levitz, Daccord, and van
  Damme~H.]{Lemaire1991}
E.~Lemaire, P.~Levitz, G.~Daccord and van Damme~H., \emph{Phys. Rev. Lett.},
  1991, \textbf{67}, 2009\relax
\mciteBstWouldAddEndPuncttrue
\mciteSetBstMidEndSepPunct{\mcitedefaultmidpunct}
{\mcitedefaultendpunct}{\mcitedefaultseppunct}\relax
\EndOfBibitem
\bibitem[Greffier \emph{et~al.}(1998)Greffier, Al~Kahwaji, and
  Kellay]{Greffier1998}
O.~Greffier, R.~J. Al~Kahwaji, A.~and and H.~Kellay, \emph{Phys. Rev. Lett.},
  1998, \textbf{91}, 3860\relax
\mciteBstWouldAddEndPuncttrue
\mciteSetBstMidEndSepPunct{\mcitedefaultmidpunct}
{\mcitedefaultendpunct}{\mcitedefaultseppunct}\relax
\EndOfBibitem
\bibitem[Ben~Salem \emph{et~al.}(2013)Ben~Salem, Cantat, and
  Dollet]{Bensalem2013}
I.~Ben~Salem, I.~Cantat and B.~Dollet, \emph{J. Fluid. Mech.}, 2013,
  \textbf{714}, 258\relax
\mciteBstWouldAddEndPuncttrue
\mciteSetBstMidEndSepPunct{\mcitedefaultmidpunct}
{\mcitedefaultendpunct}{\mcitedefaultseppunct}\relax
\EndOfBibitem
\bibitem[Zhao and Maher(1993)]{Zhao1993}
H.~Zhao and J.~V. Maher, \emph{Phys. Rev. E}, 1993, \textbf{47}, 4278\relax
\mciteBstWouldAddEndPuncttrue
\mciteSetBstMidEndSepPunct{\mcitedefaultmidpunct}
{\mcitedefaultendpunct}{\mcitedefaultseppunct}\relax
\EndOfBibitem
\bibitem[Ign\'es-Mullol \emph{et~al.}(1995)Ign\'es-Mullol, Zhao, and
  Maher]{Ignes-Mullol1995}
J.~Ign\'es-Mullol, H.~Zhao and J.~V. Maher, \emph{Phys. Rev. E}, 1995,
  \textbf{51}, 1338\relax
\mciteBstWouldAddEndPuncttrue
\mciteSetBstMidEndSepPunct{\mcitedefaultmidpunct}
{\mcitedefaultendpunct}{\mcitedefaultseppunct}\relax
\EndOfBibitem
\bibitem[Vlad \emph{et~al.}(1999)Vlad, Ign\'es-Mullol, and Maher]{Vlad1999}
D.~H. Vlad, J.~Ign\'es-Mullol and J.~V. Maher, \emph{Phys. Rev. E}, 1999,
  \textbf{60}, 4423\relax
\mciteBstWouldAddEndPuncttrue
\mciteSetBstMidEndSepPunct{\mcitedefaultmidpunct}
{\mcitedefaultendpunct}{\mcitedefaultseppunct}\relax
\EndOfBibitem
\bibitem[Chassenieux \emph{et~al.}(2011)Chassenieux, Nicolai, and
  Benyahia]{Chassenieux2011}
C.~Chassenieux, T.~Nicolai and L.~Benyahia, \emph{Curr. Opin. Colloid Interface
  Sci.}, 2011, \textbf{16}, 18\relax
\mciteBstWouldAddEndPuncttrue
\mciteSetBstMidEndSepPunct{\mcitedefaultmidpunct}
{\mcitedefaultendpunct}{\mcitedefaultseppunct}\relax
\EndOfBibitem
\bibitem[Michel \emph{et~al.}(2000)Michel, Filali, Aznar, Porte, and
  Appell]{Michel2000}
E.~Michel, M.~Filali, R.~Aznar, G.~Porte and J.~Appell, \emph{Langmuir}, 2000,
  \textbf{16}, 8702\relax
\mciteBstWouldAddEndPuncttrue
\mciteSetBstMidEndSepPunct{\mcitedefaultmidpunct}
{\mcitedefaultendpunct}{\mcitedefaultseppunct}\relax
\EndOfBibitem
\bibitem[Tabuteau \emph{et~al.}(2009)Tabuteau, Mora, Porte, Abkarian, and
  Ligoure]{Tabuteau2009}
H.~Tabuteau, S.~Mora, G.~Porte, M.~Abkarian and C.~Ligoure, \emph{Phys. Rev.
  Lett.}, 2009, \textbf{102}, 155501\relax
\mciteBstWouldAddEndPuncttrue
\mciteSetBstMidEndSepPunct{\mcitedefaultmidpunct}
{\mcitedefaultendpunct}{\mcitedefaultseppunct}\relax
\EndOfBibitem
\bibitem[Irwin(1957)]{Irwin1957}
G.~R. Irwin, \emph{Journal of Applied Mechanics}, 1957, \textbf{24}, 361\relax
\mciteBstWouldAddEndPuncttrue
\mciteSetBstMidEndSepPunct{\mcitedefaultmidpunct}
{\mcitedefaultendpunct}{\mcitedefaultseppunct}\relax
\EndOfBibitem
\bibitem[Saulnier \emph{et~al.}(2004)Saulnier, Ondar\c{c}uhu, Aradian, and
  Rapha\"{e}l]{Saulnier2004}
F.~Saulnier, T.~Ondar\c{c}uhu, A.~Aradian and E.~Rapha\"{e}l, \emph{Macromol},
  2004, \textbf{37}, 1067\relax
\mciteBstWouldAddEndPuncttrue
\mciteSetBstMidEndSepPunct{\mcitedefaultmidpunct}
{\mcitedefaultendpunct}{\mcitedefaultseppunct}\relax
\EndOfBibitem
\bibitem[Tabuteau \emph{et~al.}(2011)Tabuteau, Mora, Ciccotti, Hui, and
  Ligoure]{Tabuteau2011}
H.~Tabuteau, S.~Mora, M.~Ciccotti, C.-Y. Hui and C.~Ligoure, \emph{Soft
  Matter}, 2011, \textbf{7}, 9474\relax
\mciteBstWouldAddEndPuncttrue
\mciteSetBstMidEndSepPunct{\mcitedefaultmidpunct}
{\mcitedefaultendpunct}{\mcitedefaultseppunct}\relax
\EndOfBibitem
\bibitem[Hui \emph{et~al.}(2003)Hui, Jagota, Bennison, and Londono]{Hui2003}
C.-Y. Hui, A.~Jagota, S.~J. Bennison and J.~D. Londono, \emph{Proc. Royal Soc.
  Lond. A}, 2003, \textbf{459}, 1489\relax
\mciteBstWouldAddEndPuncttrue
\mciteSetBstMidEndSepPunct{\mcitedefaultmidpunct}
{\mcitedefaultendpunct}{\mcitedefaultseppunct}\relax
\EndOfBibitem
\bibitem[Baumberger \emph{et~al.}(2006)Baumberger, Caroli, and
  Martina]{Baumberger2006}
T.~Baumberger, C.~Caroli and D.~Martina, \emph{Eur. Phys. J. E}, 2006,
  \textbf{21}, 81\relax
\mciteBstWouldAddEndPuncttrue
\mciteSetBstMidEndSepPunct{\mcitedefaultmidpunct}
{\mcitedefaultendpunct}{\mcitedefaultseppunct}\relax
\EndOfBibitem
\bibitem[Rice(1968)]{Rice1968}
J.~R. Rice, \emph{In Fracture II, an Advanced Treatise}, Academic Press, NY,
  1968\relax
\mciteBstWouldAddEndPuncttrue
\mciteSetBstMidEndSepPunct{\mcitedefaultmidpunct}
{\mcitedefaultendpunct}{\mcitedefaultseppunct}\relax
\EndOfBibitem
\bibitem[Tixier \emph{et~al.}(2010)Tixier, Tabuteau, Carri\`ere, Ramos, and
  Ligoure]{Tixier2010}
T.~Tixier, H.~Tabuteau, A.~Carri\`ere, L.~Ramos and C.~Ligoure, \emph{Soft
  Matter}, 2010, \textbf{6}, 2699\relax
\mciteBstWouldAddEndPuncttrue
\mciteSetBstMidEndSepPunct{\mcitedefaultmidpunct}
{\mcitedefaultendpunct}{\mcitedefaultseppunct}\relax
\EndOfBibitem
\bibitem[Ramos \emph{et~al.}(2011)Ramos, Laperrousaz, Dieudonn\'{e}, and
  Ligoure]{Ramos2011}
L.~Ramos, A.~Laperrousaz, P.~Dieudonn\'{e} and C.~Ligoure, \emph{Phys. Rev.
  Lett.}, 2011, \textbf{107}, 148302\relax
\mciteBstWouldAddEndPuncttrue
\mciteSetBstMidEndSepPunct{\mcitedefaultmidpunct}
{\mcitedefaultendpunct}{\mcitedefaultseppunct}\relax
\EndOfBibitem
\end{mcitethebibliography}
\bibliographystyle{rsc}
}

\end{document}